\documentstyle[12pt,fleqn,aasms4]{article}

\newcommand{\be}{\begin{equation}}
\newcommand{\ee}{\end{equation}}
\newcommand{\bea}{\begin{eqnarray}}
\newcommand{\eea}{\end{eqnarray}}
\newcommand{\lab}[1]{\label{#1}}
\newcommand{\r}[1]{~(\ref{#1})}

\begin{document}

\title{The Hipparcos Proper Motions in Support of the Short RR Lyrae Distance Scale}
\author{Piotr Popowski\altaffilmark{1}\altaffiltext{1}{Ohio State University Presidential
Fellow} and Andrew Gould\altaffilmark{2}\altaffiltext{2}{Alfred P.\ Sloan Foundation Fellow}}
\affil{Ohio State University, Department of Astronomy, Columbus, OH 43210}
\affil{E-mail: popowski,gould@astronomy.ohio-state.edu}

\begin{abstract}

In this paper we investigate whether a misestimate of proper motions
could have been a source of substantial systematic errors in the statistical
parallax determination of the absolute magnitude of RR Lyrae stars.
In an earlier paper, we showed that the statistical parallax method is 
extremely robust and rather insensitive to various systematic effects.
The main potential problem with this method would therefore arise from
systematically bad observational inputs, primarily radial velocities and
proper motions. In that paper, we demonstrated that the radial velocities
have not been systematically misestimated. Here we turn our attention
to proper motions.
We compare three different catalogs of proper motions --- Lick, Hipparcos and
the one compiled by Wan et al.\ (WMJ). We find that the WMJ catalog
is too heterogeneous to be a reliable source.

We analyze the sample of 165 halo RR Lyrae stars with either
Lick or Hipparcos proper motions. For the stars with both Lick and Hipparcos
proper motions we use the weighted means of reported values.
Various possible biases are investigated through vigorous Monte Carlo
simulations and we evaluate small corrections due to Malmquist bias, 
anisotropic
positions of the stars on the sky, and non-Gaussian distribution of stellar
velocities.
The mean RR Lyrae absolute magnitude is $M_V=0.74\pm 0.12$ at the
mean metallicity of the sample $\left<\rm [Fe/H]\right>=-1.60$, only $0.01$ mag
brighter than the value obtained in the previous study which did not 
incorporate Hipparcos proper motions.
The faint absolute magnitudes of RR Lyrae stars confirmed by this analysis
gives strong support to the short distance scale.

\end{abstract}

\section{Introduction}
Popowski \& Gould (1998, Paper I) showed that the statistical parallax method,
which is one of the main luminosity calibration methods for RR Lyrae stars, 
is extremely robust and is insensitive to several different categories of
systematic effects.
They proved that the statistical errors are dominated
by the size of a stellar sample and that therefore future low-error 
measurements will have almost no influence on the precision of the estimate
of the absolute magnitude, unless the number of sample stars is increased.
Consequently, the main immediate avenue to improve the determination of the 
absolute magnitude of RR Lyrae stars is to eliminate measurement-related 
systematic errors.

There are three main observational pillars on which the statistical
parallax method is founded:
\begin{enumerate}
\item radial velocities,
\item proper motions,
\item dereddened apparent magnitudes.
\end{enumerate}
Substantial systematics in any of these will result in a miscalibration
of the RR Lyrae absolute magnitude. This in turn will affect an 
RR Lyrae-based distance determination to the Large Magellanic Cloud (LMC)
and so will have a crucial impact on the extragalactic distance scale.

The method presently known as ``statistical parallax'' is a combination of
secular parallax and classical statistical parallax.  
Secular parallax is based on forcing equality 
between the three first moments of the velocity distribution 
(the bulk motion $\bf w$) as determined from 
radial velocity and proper motion measurements, while classical statistical
parallax is based on forcing equality of the six second moments
(the six independent components of the velocity covariance matrix $C_{i j}$).  
In the modern combined version of statistical parallax one simply determines 
ten parameters simultaneously by applying maximum likelihood.  The ten
parameters are an overall distance scaling factor $\eta$ (relative to an 
initial arbitrary distance scale) plus the nine first and second moments,
$\bf w$ and $C_{i j}$.

In Paper I we analyzed two samples of halo RR Lyrae stars: a kinematically 
selected sample composed of 162 RR Lyrae stars taken from Layden et al. 
(1996), and a semi-independent non-kinematically selected sample of stars 
with metallicities at the [Fe/H]$\leq -1.5$. 
The later includes 106 RR Lyrae stars from Layden et al. (1996) and 
724 non-RR Lyrae stars from Beers \& Sommer-Larsen (1995).
The Beers \& Sommer-Larsen (1995) sample was used as a source of additional
radial velocities and thus allowed the investigation of possible errors
that might be caused by systematic mismeasurement of radial velocities
in the Layden et al. (1996) sample. 
The conclusion was that
radial velocities are not a likely source of a systematic error.
In this paper we turn our attention to proper motions and establish
the reliability of different proper motion catalogs.
In \S 2 we compare the Lick catalog, the Hipparcos catalog, and
the catalog compiled by Wan et al.\ (1980,WMJ) and conclude that the 
WMJ catalog is too heterogeneous to be a reliable source. We develop
uniform procedures to select data and to estimate measurement errors for 
different subclasses of stars, and we discuss some individual cases.
Finally, in \S 3 we present the result of the statistical parallax method
as applied to our new sample and the new sample supplemented by Beers \&
Sommer-Larsen (1995) radial velocities.
Our main result is to confirm  the previously established 
absolute magnitude of RR Lyrae stars, which is substantially fainter than
the estimates derived using several other methods.

\section{Proper Motion Catalogs}

There are three RR Lyrae proper motion catalogs that cover a substantial
fraction of the sky and as a result can be used in a luminosity calibration
by the statistical parallax method:
\begin{enumerate}
\item the Hipparcos catalog (ESA 1997) covering the whole sky, complete to
about V=8, but containing many stars up to about V=12,
\item the Lick NPM1 catalog covering most of the sky with declination 
$\delta > -23^{\circ}$ and covering the magnitude range between $9<{\rm V}<18$
(e.g., Klemola, Hanson \& Jones 1993),
\item the catalog compiled by Wan et al. (1980) covering most of the sky,
but with heterogeneous selection.
\end{enumerate}
In Paper I we followed Layden et al. (1996) and used the Lick Proper
Motion Catalog as our primary source. The WMJ catalog served as a 
supplementary source for RR Lyrae stars in the South.
The Hipparcos catalog, which provides precise proper motions
for bright stars in the solar neighborhood, opens an opportunity to test 
the systematics of different catalogs. As analyzed in Paper I,
the more precise proper motions of the Hipparcos catalog are not likely to
change the precision of the luminosity calibration.
Nevertheless, an inter-comparison between various catalogs is a powerful tool
for detecting hidden systematic errors or inaccurate estimates of the
measurement errors.
We begin our investigations with a total of 233 RR Lyrae stars with measured 
metallicities, apparent magnitudes, visual extinction, radial velocities,
and proper motions. The metallicities, apparent magnitudes, extinctions,
and radial velocities are taken from Layden (1994). The proper motions are 
extracted from the Hipparcos, Lick, and WMJ catalogs. 
In the initial sample consisting of both halo and disk stars, there are
12 stars with only WMJ proper motions, 20 stars with only Hipparcos proper 
motions, and 57 stars with only Lick proper motions. The proper motions
of the remaining 144 stars are measured by at least two catalogs.

\subsection{Lick NPM1 catalog}

For the stars common to both Lick and Hipparcos, we consider $\chi^2$
of the form 
$\chi^2 = \sum_i (\mu_{{\rm HIP},i}-\mu_{{\rm {\rm Lick}},i})^2/(\sigma_{{\rm HIP},i}^2+\sigma_{{\rm {\rm Lick}},i}^2)$
where $\mu_{{\rm HIP},i}$ and $\mu_{{\rm {\rm Lick}},i}$ are proper motions in either 
right-ascension
$\alpha$ or declination $\delta$ in Hipparcos and Lick, respectively, 
$\sigma_{{\rm HIP},i}$ is the Hipparcos
proper motion error as given in column 17 or 18 of the Hipparcos catalog and
$\sigma_{{\rm {\rm Lick}},i}$ is assumed to be 5 mas$\,$yr$^{-1}$ 
(e.g. Klemola et al. 1993). Excluding the three stars with measurements 
inconsistent at the level higher than 3.5$\sigma$ (where $\sigma$ is the 
combined error from Lick and Hipparcos),
we find $\chi_{\alpha}^2=69.8$ and $\chi_{\delta}^2=112.9$ for 72 stars, 
suggesting that the error in $\mu_{\delta}$ for bright stars is probably 
larger than the one
reported by Klemola et al. (1993) for fainter stars.
(Alternatively, it is possible in principle that the high value of 
$\chi_{\delta}^2$ is due to an underestimate of Hipparcos errors. However
this would imply that the Hipparcos $\mu_{\delta}$ errors were underestimated
by a factor of $\geq 3$, which we regard as extremely unlikely.)
A requirement that $\chi^2$ per degree of freedom is equal to 1 results
in the error estimate $\sigma_{{\rm {\rm Lick}},\delta}=6.7$ mas$\,$yr$^{-1}$.

\placetable{table1}

In Table 1 we list stars for which the differences between the proper motions 
in either the $\alpha$ or $\delta$ directions taken from Lick and Hipparcos 
exceed 3$\sigma$ (assuming Lick errors of 5 mas$\,$yr$^{-1}$).
The first column gives the name of a star, the second column its membership
in either the halo or disk, the third and fourth columns show the proper
motion in the $\alpha$ direction as measured by Hipparcos and Lick, 
respectively,
and sixth and seventh columns give the proper motion in the $\delta$ direction
as measured by Hipparcos and Lick, respectively. 
When the WMJ measurements are available (columns 5 and 8), they allow one to 
infer whether it is the Hipparcos or the Lick catalog that is in error.
The ninth and tenth columns give the discrepancies between Hipparcos and WMJ 
expressed in terms of the combined error $\sigma$.
For the first three stars listed in the table there is a clear indication
that the Lick proper motions are in error and that the Hipparcos proper motions
are correct. As a result we should not use the Lick
proper motions in the case of XX And. DX Del is a disk star and so does
not affect our analysis. The discrepancy of RZ Cet is in the range
between 3 and 3.5$\sigma$ and with the new suggested error of the proper 
motion in $\delta$ direction of 6.7 mas$\,$yr$^{-1}$ becomes consistent
with the assumption of the statistical fluctuation. The same is true
for the fourth star RU Cet, for which in any case there is no WMJ
measurement to resolve the discrepancy.
And finally, the star identified as ET Hya in the Hipparcos input catalog
has only a minuscule variation and its Hipparcos variability flag is empty. 
This suggests that this star was misclassified in the Hipparcos input catalog. 
Consequently, only the Lick proper motion represents an actual measurement
for this star.

Platais et al. (1998) compare Hipparcos and Lick proper motions 
for approximately 9000 stars and find
some systematics in the Lick data for stars brighter than about $V=11$. 
Stars fainter than $V=11$ do not show the same clear trend.
However, the Hipparcos errors are large for $V>11$, so systematic
effects may be present but remain undetectable.
Because the behavior of the Lick errors for stars
brighter than $V=11$ is not well understood, we decided to use
just the Hipparcos proper motion for these stars. We note, however, that using
the weighted average for all stars in common between the two catalogs
changes the final results by $<0.1\%$.
This is because if one weights Hipparcos and Lick measurements for a bright 
star based on quoted errors, the final result is dominated by the much 
more precise Hipparcos observations.

\subsection{The WMJ catalog}
A typical error of the proper motions in the Lick catalog is estimated
to be 5 mas$\,$yr$^{-1}$ (Klemola et al. 1993). Based on this 
estimate and a comparison of RR Lyrae stars
common to the Lick and WMJ catalogs, Layden et al. (1996) argue
that a typical error of the WMJ catalog is almost independent of
the error quoted by the authors and is equal to 6.5 mas$\,$yr$^{-1}$.
However, inspection of Figure 2 in Layden et al. (1996) suggests
that the WMJ stars with the highest reported errors do indeed have very
large real errors.
Moreover, it must be noted that the proper motions in the WMJ catalog consist 
only in part of new measurements, and much of the data are simply 
adopted from earlier studies.
As a result, it is possible that the typical error for the WMJ catalog as a 
whole differs from the typical error for the latitudes that overlap the Lick 
catalog.

We compare 30 stars with both Hipparcos and WMJ proper motions, but with no
Lick proper motion. That is, we test an entirely different set of stars
than the one tested by Layden et al. (1996). We
consider $\chi^2$
of the form 
$\chi^2 = \sum_i(\mu_{{\rm HIP},i}-\mu_{{\rm {\rm WMJ}},i})^2/(\sigma_{{\rm HIP},i}^2+
\sigma_{{\rm {\rm WMJ}},i}^2)$
where $\mu_{{\rm HIP},i}$ and $\mu_{{\rm {\rm WMJ}},i}$ are proper motions in 
either 
$\alpha$ or $\delta$ in Hipparcos and WMJ, respectively, 
$\sigma_{{\rm HIP},i}$ is the Hipparcos
proper motion error as given in column 17 or 18 of the Hipparcos catalog and
$\sigma_{{\rm {\rm WMJ}},i}$ is taken to be 6.5 mas$\,$yr$^{-1}$ as suggested by 
Layden et
al. (1996).
For 28 stars with measurement differences smaller than 3.5$\sigma$ (where
$\sigma$ is the combined error of Hipparcos and WMJ)
we find $\chi_{\alpha}^2=42.5$ and $\chi_{\delta}^2=51.6$, both in 
serious disagreement with the predicted value of 28. We conclude that
the WMJ catalog is very heterogeneous and its errors in the South (probed 
only by the Hipparcos catalog) must be larger than the errors in the North
(constrained by Layden et al. 1996).
We note that only 9 stars in the ``halo-3'' sample (see \S2.4) have 
exclusively WMJ measurements. We will ultimately remove these 9 stars from our
analysis and base our study entirely on the Lick and Hipparcos catalogs for
which we have a much better understanding of the errors.

\placetable{table2}

In Table 2 we list stars for which the proper motions in either the $\alpha$
or $\delta$ direction taken
from WMJ and Hipparcos differ by more than 3$\sigma$ (assuming WMJ error
of 6.5 mas$\,$yr$^{-1}$). 
The first column gives the name of a star, the second column its membership
in either the halo or disk, the third and fourth columns show the proper
motion in $\alpha$ direction as measured by Hipparcos and WMJ, respectively,
the fifth and sixth columns give the proper motion in $\delta$ direction
as measured by Hipparcos and WMJ, respectively, and the seventh and eighth
columns give the discrepancies between Hipparcos and WMJ as expressed in terms
of the combined error $\sigma$. 
There is no clear indication which of the surveys is in error.
The discrepancies of the first two stars: AT And and V675 Sgr are so large that
they are not likely to result from Gaussian or semi-Gaussian process and so 
we exclude these stars from further analysis.
With the increased error of the WMJ proper motions suggested by our $\chi^2$ 
analysis, the measurements of the remaining two stars are consistent with the 
hypothesis that the Hipparcos values are correct. We therefore include them.

\subsection{Comparison between Lick and WMJ catalogs}

Here we conduct an analysis of the proper motions of RR Lyrae stars that are 
common to Lick and WMJ catalogs. We summarize our findings in Table 3.

\placetable{table3}

The first column gives the name of a star, the second column its membership
in either the halo or disk, the third and fourth columns show the proper
motion in the $\alpha$ direction as measured by Lick and WMJ, respectively,
and the fifth column gives the proper motion error in the $\alpha$ direction
as quoted by WMJ.
Columns 6, 7 and 8 list the proper motion in the $\delta$ direction
as measured by Lick and WMJ, respectively, and the proper motion error in 
$\delta$ direction as quoted by WMJ.
The first two stars were measured by Hipparcos.
The Hipparcos proper motion of VY Ser is $(\mu_{\alpha,{\rm HIP}},\mu_{\delta,{\rm HIP}})= (-103.2,-10.1)\, {\rm mas\,yr^{-1}}$ in good agreement with the 
Lick value. Note that WMJ assigned very generous errors to this star. This case
confirms that there is a strong correlation between the precision of
the measurement and error quoted by WMJ for the stars with the largest 
quoted errors.
The Hipparcos proper motion of BB Vir is $(\mu_{\alpha,{\rm HIP}},\mu_{\delta,{\rm HIP}})= (-49.7,-3.5)\, {\rm mas\,yr^{-1}}$, which is closer to the WMJ 
value in the $\alpha$ direction and closer to the Lick value in 
the $\delta$ direction.
In this case, where WMJ's quoted error is small, the Hipparcos measurement
confirms that the true WMJ error is in the normal statistical range.
Encouraged by our experience with VY Ser and BB Vir, when lacking Hipparcos
measurements, we will judge the significance of
the discrepancy of Lick and WMJ measurements by the size of the error
quoted by WMJ.
When the error is very large (20-30 ${\rm mas\,yr^{-1}}$) we assume that
there is no information in the WMJ value, when the error is 
~10$\,{\rm mas\,yr^{-1}}$ or less we assume that the error is 
~8-9$\,{\rm mas\,yr^{-1}}$ regardless of the quoted value.
We apply this approach to several suspicious stars and identify four
stars with significant disagreements. They are listed at the bottom of Table 3.
BB Pup is a disk star and so will not affect our determination.
We remove the remaining three stars from the further analysis.
One might argue that RY Com should be included, because its huge
error in the $\alpha$ direction $\epsilon_{\alpha,{\rm WMJ}}=23.2 \,
{\rm mas\,yr^{-1}}$ suggests that WMJ
measurements are highly unreliable. However the reported error in $\delta$ 
direction is only  $\epsilon_{\delta,{\rm WMJ}}=2.9 \, {\rm mas\,yr^{-1}}$ and so
it is not obvious
that the WMJ measurement should be ignored.
We prefer to be cautious and therefore remove this star.
 
\subsection{The Final Data Set}

Out of 233 stars in our initial sample we first select 177 stars that we 
define as halo. This sample of halo stars corresponds to halo-3 stars
defined by Layden et al. (1996), i.e. stars that meet the following criteria:
\begin{enumerate}
\item all stars with $V_{\theta}< [-400 {\rm [Fe/H]} - 300]\,{\rm km\,s^{-1}}$,
\item but excluding stars with $|V_{\pi}|<100 \,{\rm km\,s^{-1}}$, 
$|V_{\theta}|>80 \,{\rm km\,s^{-1}}$, $|V_z|<60 \,{\rm km\,s^{-1}}$, $|z|<1\,{\rm kpc}$ and ${\rm [Fe/H]}>-1.6$.
\end{enumerate}
Among the stars selected this way, 9 lack both Lick and Hipparcos proper 
motions and so are excluded from further consideration.
We make the following adjustments based on information about individual
stars which yield our final sample of 165 stars.
\begin{description}
\item[ET Hya:] In the Hipparcos catalog this star has only a minuscule 
variation and its variability flag is empty, suggesting that this star was 
misclassified in the input Hipparcos catalog. Consequently,
we use only the Lick proper motion instead of weighting Lick and Hipparcos
proper motions.
\item[AT And:] A cross-check between Hipparcos and WMJ shows a major discrepancy
($\mu_{\delta,{\rm HIP}}=-50.3\, {\rm mas\,yr^{-1}}$ and $\mu_{\delta,{\rm WMJ}}=46.0 
\,{\rm mas\,yr^{-1}}$).
WMJ do not quote a large error for this star so there is no reason to give
preference to either measurement. The star is removed from further 
analysis.
\item[V675 Sgr:] A cross-check between Hipparcos and WMJ shows a major 
discrepancy
[$(\mu_{\alpha,{\rm HIP}},\mu_{\delta,{\rm HIP}})=(30.9,-22.0)\, {\rm mas\,yr^{-1}}$ and
$(\mu_{\alpha,{\rm WMJ}},\mu_{\delta,{\rm WMJ}})=(0.0,12.0)\, {\rm mas\,yr^{-1}}$].
WMJ do not quote a large error for this star so there is no reason to give
preference to either measurement. The star is removed from further 
analysis.
\item[RX CVn:] A cross-check between Lick and WMJ shows a major 
discrepancy
[$(\mu_{\alpha,{\rm Lick}},\mu_{\delta,{\rm Lick}})=(-2.0,1.0)\, {\rm mas\,yr^{-1}}$ and
$(\mu_{\alpha,{\rm WMJ}},\mu_{\delta,{\rm WMJ}})=(49.8,-27.3)\, {\rm mas\,yr^{-1}}$].
WMJ do not quote a particularly large error for this star so there is no 
reason to give preference to either measurement. The star is removed from 
further analysis.
\item[RY Com:] A cross-check between Lick and WMJ shows a major 
discrepancy
[$(\mu_{\alpha,{\rm Lick}},\mu_{\delta,{\rm Lick}})=(-6.1,-17.8)\, {\rm mas\,yr^{-1}}$ and
$(\mu_{\alpha,{\rm WMJ}},\mu_{\delta,{\rm WMJ}})=(-16.6,13.5)\, {\rm mas\,yr^{-1}}$].
WMJ quote a large error only for the less discrepant component of this star's 
proper motion but there is no compelling case favoring the Lick measurement
for the more discrepant component. 
The star is removed from further analysis.
\item[Z Com:] A cross-check between Lick and WMJ shows a major 
discrepancy
[$(\mu_{\alpha,{\rm Lick}},\mu_{\delta,{\rm Lick}})=(-7.7,-18.5)\, {\rm mas\,yr^{-1}}$ and
$(\mu_{\alpha,{\rm WMJ}},\mu_{\delta,{\rm WMJ}})=(15.0,-3.0)\, {\rm mas\,yr^{-1}}$].
WMJ do not quote a particularly large error for this star so there is no 
reason to give preference to either measurement. The star is removed from 
further analysis.
\item[AO Peg:] Following Layden et al. (1996) we reclassify this star as a halo
member due to its extreme $\pi$ and $z$ velocity components.
\item[FU Vir:] Following Layden et al. (1996) we reclassify this star as a halo
member due to its extreme $\pi$ and $z$ velocity components.
\end{description}
Table 1 shows that for XX And the Lick catalog was in error and, as a result, 
one should use only the Hipparcos proper motion.
We did not include XX And in the list above because its absolute $V$ magnitude
is 10.56 and so according to our general rule for stars brighter than $V=11$,
we use the Hipparcos proper motion for this star anyway.

It is important to realize that eliminating outliers and suspicious stars
from the original data set is very unlikely to introduce any bias
to the solution. Our data selection procedure is selection a priori and
is based on the comparison of different sources of data and
not on the results of the statistical parallax analysis.
Only the membership change for AO Peg and FU Vir relies in some way
on the results of the analysis, but even their reassignment may
be interpreted as only a small change in the classification definition.

Among the stars that enter our final sample 47 have only Lick proper motions
and 16 only Hipparcos proper motions. That is, 102 stars out of 165 
(62\% of the sample) have been cross-checked for non-statistical errors
between at least two different catalogs.

\section{Results}

We apply the equations derived in Paper I and Newton's method to find 
the maximum likelihood solution for our sample of halo RR 
Lyrae stars.
We conduct vigorous Monte Carlo simulations to correct for various
systematic effects. The reader is referred to Paper I which describes our
method of Monte Carlo investigations in detail.
In each simulation we construct 4000 mock samples of 165 halo stars
(165 is the size of our final sample).
For each sample we generate a set of 165 space velocities drawn
from a distribution with specified means, dispersions and kurtoses
in the three principal directions.
For each star we transform its velocity components from the star's Galactic 
frame of reference to the Sun's frame and add Gaussian measurement errors.
In the second step we find the most probable
parameters describing each of the samples. To analyze our mock
samples we use exactly the same maximum likelihood procedure that is used 
to obtain the results for the real stars.

Table 4 compares the basic new results with the ones obtained
in the previous analysis.
We work in terms of a distance scale that is defined in such a way that
if our determination were in perfect agreement with Layden et 
al. (1996), the scaling parameter $\eta$ would be exactly equal to 1.
The three rows give the maximum likelihood
values of the ten parameters characterizing different samples.
All results were corrected for numerical and positional biases.
We also take into account
that maximum likelihood underestimates the velocity variances by a factor 
$N/(N-1)$.
The direct effects of the possible scatter of the absolute magnitude of
RR Lyrae stars are estimated
numerically by introducing artificial scatter into the Monte Carlo simulations.
The indirect effects of Malmquist (1920) bias are as described analytically
in Paper I.

\placetable{table4}

The first row repeats the result obtained in Paper I.
The next two rows are based on the data described in detail in
\S 2. The second row contains the results obtained assuming proper motion 
errors in the $\delta$ direction of 6.7 mas$\,$yr$^{-1}$.
The third row assumes proper motion errors in the 
$\delta$ direction of 5.0 mas$\,$yr$^{-1}$.
This change in the assumption about the proper motion errors alters 
$\eta$ by less than 1\%.

We assume that RR Lyrae stars follow the absolute magnitude-metallicity 
relation 
($M_{V}= {\rm const} + 0.15 {\rm [Fe/H]}$) of Carney, Storm, \& Jones (1992).
However, the results are only sensitive to the value of the absolute magnitude
at the mean metallicity of the sample, $\left< {\rm [Fe/H]} \right> = -1.60$, 
and not to the slope 
of the relation. We checked that the solutions for different slopes are 
statistically indistinguishable from one another.
By taking account of the metallicity, we restrict the possible scatter in 
RR Lyrae absolute magnitudes to the intrinsic scatter at fixed metallicity.
To correct for the scatter in the absolute magnitude, we adopt 
${\sigma}_M=0.15$.
To make this estimate, we first inspected color-magnitude diagrams of
several globular clusters and found that for these {\em very homogeneous}
populations the dispersion is typically ${\sigma}_M=0.08$. We take this
as a lower limit. The only available {\em inhomogeneous} sample at
approximately fixed distance is the 28 RRab Lyrae stars of Hazen \& Nemec 
(1992), which are 4$^{\circ}$ east of the LMC center and for which 
${\sigma}_M \sim 0.17$. We take this as an upper limit because some
of the scatter may be due to dispersion in distance. For example,
if the LMC RR Lyrae population has an $r^{-3.5}$ profile and a flattening
$c/a=0.6$ (similar to the Galactic population), we find a dispersion
due to distance ${\sigma}_{M,dist}=0.09$, implying an intrinsic dispersion
of ${\sigma}_M \sim 0.14$.
In any event, those prefering other values of ${\sigma}_M$ should note that 
it is straightforward to find the corrected values of $\eta$ using
equation\r{scatt.cor}

The correction we apply in constructing Table 4 is
\be
\frac{\delta \eta}{\eta} \approx 0.25 {\sigma}_M^2 - 0.64 {\sigma}_M^2 = -0.39 {\sigma}_M^2. \lab{scatt.cor}
\ee
where the first term, estimated both analytically and numerically, is due to 
the scatter in the 
absolute magnitude of RR Lyrae stars (\S 3.1 of Paper I) and the second term 
is due to Malmquist bias (\S 3.2 of Paper I).

We use the following notation:
\be
\sigma_{\pi}\equiv C_{11}^{\frac{1}{2}}, \;\;\;\;\;\;\;\;\;\;\sigma_{\theta} \equiv C_{22}^{\frac{1}{2}}, \;\;\;\;\;\;\;\;\;\;\sigma_z\equiv C_{33}^{\frac{1}{2}},
\ee
stressing that we are
measuring the underlying velocity distribution in the {\it local} 
Galactic frames of the stars in the sample under the assumption of 
Galactic axisymmetry. 
Our best estimate of the RR Lyrae absolute magnitude at the mean metallicity 
of the sample $\left<\rm [Fe/H]\right> =-1.60$ is $M_V=0.74\pm 0.12$.
The velocity ellipsoid is $(\sigma_{\pi},\sigma_{\theta},\sigma_z)=(170\pm 10, 97\pm 7,
93\pm 7)\,\, \rm{km} \, {\rm s}^{-1}$ and the RR Lyrae population is moving
in $\theta$ direction at $-214 \pm 11 \,\, \rm{km} \, {\rm s}^{-1}$ relative
to the Sun, where the errors are taken from the second row of the Table 4.
Additionally we report our new best estimates for the kurtoses of the RR Lyrae 
velocity distribution in the three principal directions:
$K_{\pi}=1.95$, $K_{\theta}=3.25$, and $K_{z}=4.50$. The new results confirm
the highly non-Gaussian character of the RR Lyrae velocity distribution
(see also Figure 1 in Paper I).

We also analyze a non-kinematically selected sample of stars with
$\rm [Fe/H] \leq -1.5$ taken from Beers \& Sommer-Larsen (1995) and 
Layden (1994), supplemented by Hipparcos proper motions.
The new combined sample consists of a total of 827 stars: 
724 non-RR Lyrae stars come from the  Beers \& Sommer-Larsen (1995) sample 
and 103 
RR Lyrae stars 
\footnote{About 10\% of these 103 stars are 
different than in Paper I: we added a few new stars present in Layden (1994)
with Hipparcos proper motions but
not in Layden et al. (1996), and we removed a few e.g., those with only WMJ 
proper motions.} from Layden (1994).
The mean metallicity of 103 RR Lyrae stars is $\left<\rm [Fe/H]\right> = 
-1.79$, and the mean
metallicity of the stars taken from Beers \& Sommer-Larsen (1995) sample
is $\left<\rm [Fe/H]\right> = -2.21$. The conclusions of the following analysis are valid only as
long as kinematics of halo stars do not depend critically on the metallicity
(in the metallicity range spanned by the two groups of stars).
Beers \& Sommer-Larsen (1995) find that kinematics do not vary significantly
with metallicity in this range. However, if there were a correlation
below the limit of their detection, the expected sign would be more
extreme kinematics at lower metallicity. Such an effect, if real, would cause
us to {\it overestimate} the RR Lyrae luminosity, and its correction
would therefore make the RR Lyrae distance scale even shorter.

\placetable{table5}

In Table 5 we compare the old and new parameter and error determinations for 
the combined samples.
The results are corrected for magnitude dispersion, Malmquist bias,
and other effects mentioned above.
The Malmquist bias correction for the non-kinematic sample is the 
same as the one applied in Table 4.  The magnitude dispersion
correction was determined based on numerical simulations. 
Consequently, we compare final results: 
the best values and uncertainties for the old combined sample
are repeated from Paper I, and the corresponding new data for two different
error estimates in the $\delta$ component of the proper motion
are presented in the next two rows (second row with 6.7 mas$\,$yr$^{-1}$ error 
and third row with 5.0 mas$\,$yr$^{-1}$ error).
We conclude that RR Lyrae absolute magnitude at the mean metallicity 
of the sample $\left<\rm [Fe/H]\right> =-1.79$ is $M_V=0.80\pm 0.11$.
The velocity ellipsoid is $(\sigma_{\pi},\sigma_{\theta},\sigma_z)=(161\pm 6, 108\pm 8,
94\pm 5)\,\, \rm{km} \, {\rm s}^{-1}$ and the stars are moving
in $\theta$ direction at $-198 \pm 8 \,\, \rm{km} \, {\rm s}^{-1}$ relative
to the Sun, where the errors are taken from second row of the Table 5.
The three results are very close to one another, just as was the case
for the corresponding results in Table 4.

\section{Conclusions}

This paper was in part motivated by the recent {\it Letter} by Tsujimoto 
et al. (1998) who analyzed a smaller RR Lyrae sample for which there exist 
Hipparcos measurements. The central value of absolute magnitude obtained by 
Tsujimoto et al., $M_V = 0.69 \pm 0.10$ at the average metallicity of 
$\left<\rm [Fe/H]\right> = -1.58$, 
is in relatively 
good agreement with the value derived here, but the two error estimates
are in striking disagreement. We would like to stress that the error 
intrinsic to statistical parallax is completely dominated by 
sample-size effects and that the high precision of the Hipparcos 
proper motions allows only a marginal improvement in the precision of the 
absolute magnitude measurement.
The errors quoted by Tsujimoto et al. (1998) for 99 stars are substantially 
smaller than the theoretical minimum which we derived analytically in Paper I.
(We also attempted to confirm Tsujimoto et al.'s (1998) 
finding of a significant rotation between the Lick and Hipparcos frames.
However, we found instead that the rotation is consistent with zero.)

The apparent magnitudes of RR Lyrae stars remain the only major 
unchecked-for-systematics ingredient of the statistical parallax analysis.
In appendix A, we describe the results of our (not very conclusive) attempt to
address this issue by comparing Hipparcos and Layden (1994) apparent
magnitudes. Our analysis suggests a possible correction toward fainter values
of the absolute magnitudes of the field RR Lyrae stars, not in the right
direction to reconcile the discrepant distance scales.

In Paper I we showed that systematic problems with radial velocities
are not likely to cause a miscalibration of the RR Lyrae absolute magnitude.
Here we have shown that systematic problems with proper motions are also
not likely to cause a miscalibration. 
Because the statistical parallax method is very insensitive to other
systematic effects, we conclude that the statistical error
of 0.12 mag reported above for the absolute magnitudes of field RR Lyrae
stars is the true error. The statistical parallax calibration
$M_V=0.74\pm 0.12$ at $\left<\rm [Fe/H]\right> = -1.60$ is therefore inconsistent at the 2$\sigma$
level with that derived from main-sequence fitting by Reid (1997) ($ M_V \sim 
0.44 \pm 0.07$ at [Fe/H]=$-1.6$) and Gratton et al. (1997) 
($ M_V = 0.47 \pm 0.04$ at [Fe/H]=$-1.6$) using Hipparcos parallaxes of 
nearby subdwarfs.
That is, Hipparcos proper motions of RR Lyrae stars confirm a short local
distance scale.

Gratton (1998) suggested that there may be a difference in the luminosity
between globular cluster and field RR Lyrae stars.
If RR Lyrae stars in globular cluster belonged
to different population the problem would certainly be solved, but
Catelan (1998) showed that RR Lyrae stars in these two different environments
have essentially the same distribution in the period-equilibrium temperature
plane, both at the metal-poor and at the metal-rich ends, suggesting
very similar luminosities.
He also argued that it is difficult to adjust helium abundance of RR Lyrae 
stars in different environments in such a way as to produce the required 
luminosity difference of 0.2 mag and still maintain the striking overlap 
in the period-equilibrium temperature plane.
It is however conceivable that some characteristics of RR Lyrae stars
may be sensitive to the presence (globular clusters) or absence (field) of
non-canonical deep mixing that occurs during the red giant phase 
(Kraft et al. 1997).

\acknowledgements
We are very grateful to Robert Hanson for providing information about Lick NPM1
catalog and a thorough discussion of some individual stellar measurements.
We thank Takuji Tsujimoto for sending us part of the data prior to publication
of his ApJ {\it Letter}.
We are in debt to Michele Turansick for copying for us the paper by Wan et al.
(1980) which was not available in our library.
We also would like to thank Imants Platais for sending us his paper prior to 
its publication.
This work was supported in part by grant AST 94-20764 from the NSF.

\clearpage

\appendix
\section{Comparison of Apparent Magnitudes}

Here we investigate whether apparent magnitudes of RR Lyrae stars taken
from two different sources, the Hipparcos catalog and Layden (1994), are 
consistent with each other.
The Variability Annex to the Hipparcos catalog
gives the faintest and brightest values of apparent V magnitudes for variable
stars, while Layden (1994) gives V magnitude at mean intensity $\langle V 
\rangle_I$.
To compare the apparent magnitudes, we apply two approximate
relations (e.g., Strugnell et al. 1986),
$$ \langle V \rangle_I = 0.5(V_{faintest}+V_{brightest}) + 0.07, $$
$$ \langle V \rangle_I = 0.61 V_{faintest} - 0.39 V_{brightest} -0.05, $$
to the Hipparcos measurements. Then we plot the difference between Layden 
(1994) and 
Hipparcos-based
apparent magnitudes as a function of Layden (1994) apparent magnitude.
There is no obvious trend of this difference with increasing magnitude.
However, the scatter in the difference increases dramatically as the Hipparcos
magnitude limit, $V \sim 12$, is approached. We checked the apparent 
magnitudes of 
the most discrepant cases with the SIMBAD database and found out that, as 
expected, the large disagreement is almost always caused by an Hipparcos 
error.
The most interesting result of this exercise is that
the average difference between apparent magnitudes is not consistent with
zero. The Hipparcos magnitudes are on average 0.1 mag fainter than the 
Layden (1994) magnitudes. This suggests that either the photometry of 
Hipparcos is too faint or that the photometry of Layden (1994) is too 
bright or both.
The results quoted in the main part of the paper are based on Layden (1994)
magnitudes. If Hipparcos magnitudes are correct, then the absolute magnitude
of RR Lyrae stars is $0.1$ mag fainter than indicated by our analysis
and in even more serious disagreement with new results from globular clusters.

\clearpage

\begin{table}
\dummytable\label{table1}
\end{table}

\begin{table}
\dummytable\label{table2}
\end{table}

\begin{deluxetable}{cccccccc}
\tablewidth{0pt}
\tablecaption{Lick vs. WMJ --- data for discrepant stars. \label{table3}}
\tablehead{
\colhead{Star name} & \colhead{Membership} &\colhead{$\mu_{\alpha,{\rm Lick}}$} & \colhead{$\mu_{\alpha,{\rm WMJ}}$} & \colhead{$\epsilon_{\alpha,{\rm WMJ}}$} & \colhead{$\mu_{\delta,{\rm Lick}}$} & \colhead{$\mu_{\delta,{\rm WMJ}}$} & \colhead{$\epsilon_{\delta,{\rm WMJ}}$}
}
\startdata
VY Ser & halo & -98.4\phm{1} & -58.8\phm{1} & 32.6\phm{1} & -5.1 & 29.4\phm{1} & 21.5\phm{1} \nl
BB Vir & halo & -37.1\phm{1} & -44.4\phm{1} & 0.9 & -10.2\phm{1} & 10.4\phm{1} & 4.8 \nl
&&&&&&&\nl
RX CVn & halo & -2.0 & \phm{-}49.8\phm{1} & 11.4\phm{1} & \phm{-}1.0 & -27.3\phm{-1} & 0.4 \nl
RY Com & halo & -6.1 & -16.6\phm{1} & 23.2\phm{1} & -17.8\phm{1} & 13.5\phm{1} & 2.9 \nl
BB Pup & disk & -15.8\phm{1} & \phm{-}7.0 & 7.0 & \phm{-}10.5\phm{1} & 6.0 & 7.0 \nl
Z Com & halo & -7.7 & \phm{-}15.0\phm{1} & 5.0 & -18.5\phm{1} & -3.0\phm{-} & 5.0 \nl
\enddata
\tablecomments{All proper motions are expressed in units of ${\rm mas \, 
yr^{-1}}$. $\epsilon_{\alpha,{\rm WMJ}}$ and $\epsilon_{\delta,{\rm WMJ}}$ are the errors 
taken from WMJ.}
\end{deluxetable}

\begin{table}
\dummytable\label{table4}
\end{table}

\begin{table}
\dummytable\label{table5}
\end{table}

\end{document}